# Weighing $\Delta E = c^2 \Delta m$ on a relativistic high-sensitivity balance. Its mechanical and thermal equilibrium.

Bernhard Rothenstein[1], Stefan Popescu[2] and George J. Spix[3]

1) Politehnica University of Timisoara, Physics Department, Timisoara, Romania
2) Siemens AG, Erlangen, Germany
3) BSEE Illinois Institute of Technology, USA

**Abstract.** *We consider the sticky collision between two sphere of equal rest mass, moving with equal but opposite speeds in a horizontal plane. The collision takes place on the pan of a high sensitivity balance. The conditions of mechanical and thermal equilibrium are studied showing that heat developed in the rest frame of the system has a corresponding inertial mass.*

The origin and the meaning of Einstein's equation $\Delta E = c^2 \Delta m$ can be synthesised by stating that **mass and energy are different manifestations of the same thing, the energy having mass and the mass being a form of energy.** Taking into account that the Einstein's equation about the equivalence of mass and energy is often misinterpreted in popular approaches, we make use of thought experiments that could illuminate the way in which we should understand it.[1]

A thought experiment is a technique for conceptual reasoning. It could be used to convey the essence of the basic anti-common sense associated with special relativity helping us to derive the basic formulas of special relativity like time dilation and length contraction. The thought experiment could be also used in order to find out the result of an experiment using the formulas offered by theory, involving in many cases imaginary measuring devices whose sensitivity excesses the sensitivity of the real ones.

Einstein's formula $\Delta E = c^2 \Delta m$ establishes a relationship between the change in the energy $\Delta E$ and the change in mass of the same physical system. It can be derived in many different ways, even without involving special relativity.[2]

The thought experiment we propose involves a high sensitivity balance with stationary pans as shown in Figure 1. Consider on its left pan two identical spheres of rest mass $m_0$ moving in the horizontal plane with equal and opposite speeds $V$ and $-V$ respectively. Let $T_0$ be the temperature of the ambient within which the spheres are in a state of thermal equilibrium with the ambient. The problem of the sticky collision between the two spheres has two aspects. Before the collision the two spheres are in a state of thermal equilibrium with the ambient and in a state of mechanical equilibrium if we place on the right pan a body of rest mass $M_0$ the energy of which should equate the total energy of the two spheres moving on the left pan i.e.

$$\frac{2m_0 c^2}{\sqrt{1-\frac{V^2}{c^2}}} = M_0 c^2 \qquad (1)$$

the conditions of mechanical and thermal equilibrium being fulfilled.

After the perfectly inelastic collision the two spheres coalesce into a single one at rest. The initial kinetic energy is converted into heat and as a result the temperature of the compound particle increases becoming equal to $T>T_0$. Under such conditions the balance is still in a state of mechanical equilibrium but out of thermal equilibrium as a result of the fact that its temperature has increased. **The amount of heat developed in the rest frame has a corresponding inertial mass.**[3]

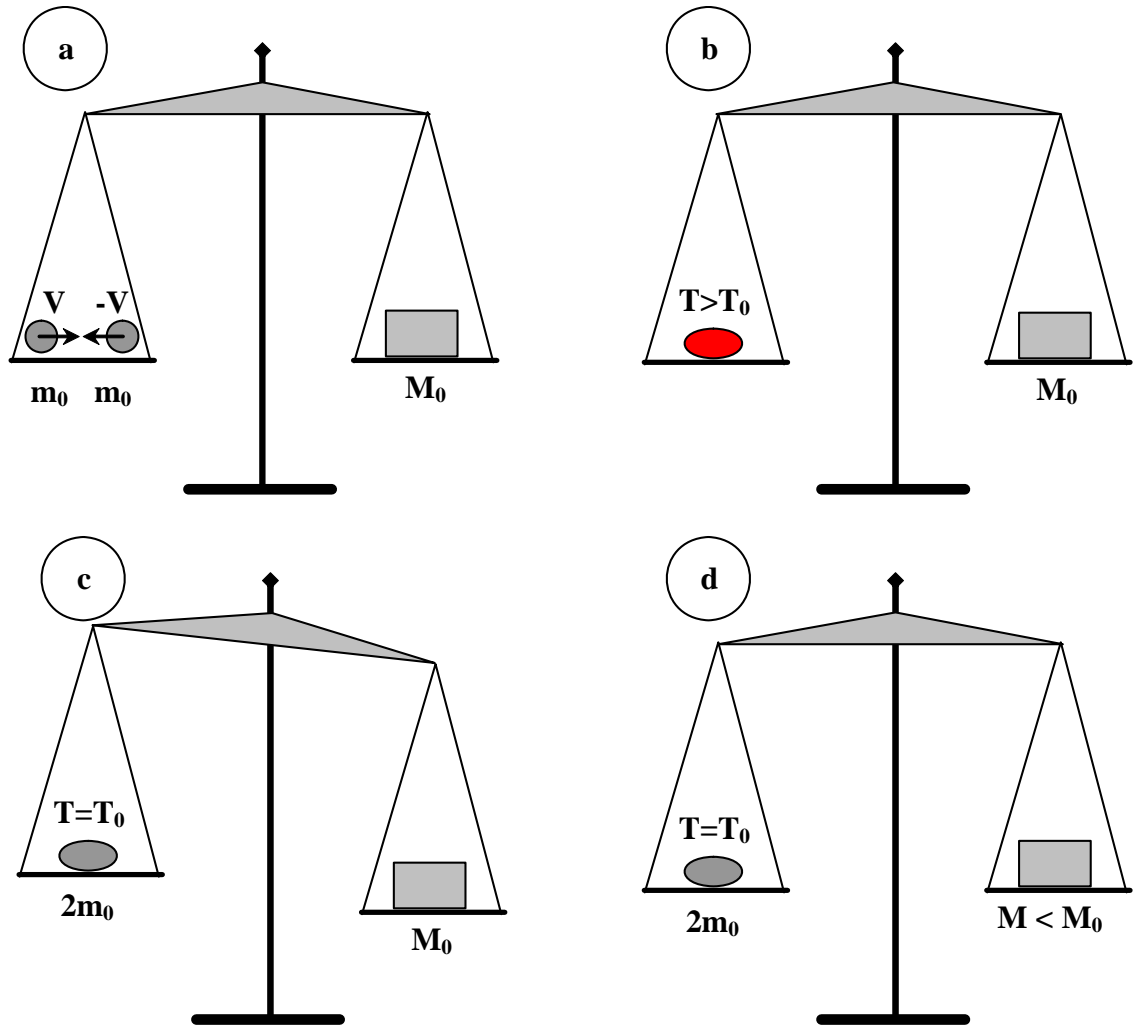

**Figure 1**. Weighing $\Delta E = c^2 \Delta m$ on a relativistic high-sensitivity balance.
  a. The balance in mechanical and thermal equilibrium before the sticky collision.
  b. The balance immediately after the sticky collision in mechanical equilibrium but not in thermal equilibrium.
  c. The balance after the radiation of the heat. The balance is in a state of thermal equilibrium but not in mechanical equilibrium.
  d. The balance in thermal in mechanical equilibrium

The compound sphere radiates the heat until its temperature becomes equal to $T_0$ when the balance is in a state of thermal equilibrium but not in a state of mechanical equilibrium. Removing from the right pan a mass

$$\Delta m = \frac{E_k}{c^2} = 2m_0 c \left[ \frac{1}{\sqrt{1 - \frac{V^2}{c^2}}} - 1 \right] \quad (2)$$

(6)

that corresponds to the initial kinetic energy the mechanical equilibrium is restored, the final mass on the right pan becoming as expected

$$M = M_0 - \Delta m = 2m_0 \quad (3)$$

and the system is now in a state of thermal and mechanical equilibrium.

We consider that the thought experiment presented above illustrates the correct way in which we should apply Einstein's celebrated equation. It also shows that a thought experiment proposed to students could test if they have a correct representation about $\Delta E = c^2 \Delta m$ and about the heat-inertial mass equivalence.